\documentclass[12pt]{article}
\usepackage{amsmath,amsfonts,amssymb,latexsym}
\usepackage{pdfsync}
\usepackage{calrsfs}

\newtheorem{theorem}{\textbf{Theorem}}
\newtheorem{cor}[theorem]{\textbf{Corollary}}
\DeclareMathSymbol{\Gamma}{\mathalpha}{letters}{"00}
\DeclareMathSymbol{\Delta}{\mathalpha}{letters}{"01}
\DeclareMathSymbol{\Theta}{\mathalpha}{letters}{"02}
\DeclareMathSymbol{\Lambda}{\mathalpha}{letters}{"03}
\DeclareMathSymbol{\Xi}{\mathalpha}{letters}{"04}
\DeclareMathSymbol{\Pi}{\mathalpha}{letters}{"05}
\DeclareMathSymbol{\Sigma}{\mathalpha}{letters}{"06}
\DeclareMathSymbol{\Upsilon}{\mathalpha}{letters}{"07}
\DeclareMathSymbol{\Phi}{\mathalpha}{letters}{"08}
\DeclareMathSymbol{\Psi}{\mathalpha}{letters}{"09}

\DeclareMathSymbol{\Omega}{\mathalpha}{letters}{"0A}

\DeclareMathOperator{\Div}{Div}
\DeclareMathOperator{\mdiv}{div}

\DeclareMathOperator{\tr}{tr}

\newcommand{\nc}{\newcommand}

\nc{\rs}{\mathcal}
\nc{\R}{\mathbb{R}}
\nc{\N}{\mathbb{N}}
\nc{\Q}{\mathbb{Q}}
\nc{\ST}{\mathbb{S}}
\nc{\C}{\mathbb{C}}
\nc{\Z}{\mathbb{Z}}
\nc{\oH}{\mathaccent'27H}
\nc{\oW}{\mathaccent'27W}
\nc{\obH}{\mathord{\setbox0=\hbox{$\boldsymbol{\mathaccent'27H}$}\setbox1=\hbox{$\boldsymbol H$}\ht0=\ht1\box0}}
\nc{\lap}{\triangle}

 \nc{\cA}{\rs{A}}
 \nc{\cB}{\rs{B}}
\nc{\cC}{\rs{C}}
\nc{\cD}{\rs{D}}
\nc{\cG}{\rs{G}}
\nc{\cE}{\rs{E}}
\nc{\cF}{\rs{F}}
\nc{\cH}{\rs{H}}
\nc{\cK}{\rs{K}}
\nc{\cM}{\rs{M}}
\nc{\cP}{\rs{P}}
\nc{\cS}{\rs{S}}
\nc{\cT}{\rs{T}}
\nc{\cU}{\rs{U}}
\nc{\cV}{\rs{V}}
\nc{\cX}{\rs{X}}
\nc{\cY}{\rs{Y}}

\nc{\gA}{\mathfrak A}
\nc{\gF}{\mathfrak{F}}
\nc{\gG}{\mathfrak{G}}
\nc{\gJ}{\mathfrak J}
\nc{\gH}{\mathfrak {H}}

\nc{\gS}{\mathfrak S}
\nc{\gX}{\mathfrak X}
\nc{\gY}{\mathfrak Y}

\nc{\Om}{\Omega}
\nc{\Ga}{\Gamma}
\nc{\vs}{\varSigma}
\nc{\vf}{\varPhi}
\nc{\la}{\langle}
\nc{\ra}{\rangle}
\nc{\De}{\Delta}
\nc{\de}{\delta}
\nc{\ti}{\widetilde}
\nc{\wh}{\widehat}
\nc{\ov}{\overline}
\nc{\wto}{\rightharpoonup}
\nc{\emb}{\hookrightarrow}

\nc{\fu}{\mathfrak{u}}
\nc{\ba}{\boldsymbol{a}}
\nc{\bb}{\boldsymbol{b}}
\nc{\be}{\boldsymbol{e}}
\nc{\fb}{\boldsymbol{f}}
\nc{\bg}{\boldsymbol{g}}
\nc{\bh}{\boldsymbol{h}}
\nc{\bell}{\boldsymbol{\ell}}
\nc{\bn}{\boldsymbol{n}}
\nc{\bu}{\boldsymbol{u}}
\nc{\bv}{\boldsymbol{v}}
\nc{\bw}{\boldsymbol{w}}
\nc{\bx}{\boldsymbol{x}}
\nc{\by}{\boldsymbol{y}}
\nc{\bz}{\boldsymbol{z}}
\nc{\bF}{\boldsymbol{F}}
\nc{\bL}{\boldsymbol{L}}
\nc{\bH}{\boldsymbol{H}}
\nc{\bM}{\boldsymbol{M}}
\nc{\bU}{\boldsymbol{U}}
\nc{\bY}{\boldsymbol{Y}}
\nc{\bB}{\boldsymbol{B}}
\nc{\bbS}{\mathbb{S}}
\nc{\bX}{\boldsymbol{X}}
\nc{\btau}{\boldsymbol{\tau}}
\nc{\bsigma}{\boldsymbol{\sigma}}
\nc{\btheta}{\boldsymbol{\theta}}
\nc{\bzeta}{\boldsymbol{\zeta}}
\nc{\bvarphi}{\boldsymbol{\varphi}}
\nc{\bph}{\boldsymbol{\varphi}}
\nc{\bnu}{\boldsymbol{\nu}}
\nc{\bxi}{\boldsymbol{\xi}}
\nc{\beeta}{\boldsymbol{\eta}}
\nc{\bbeta}{\boldsymbol{\beta}}
\nc{\blambda}{\boldsymbol{\lambda}}
\nc{\bvarepsilon}{\boldsymbol{\varepsilon}}
\nc{\bvep}{\boldsymbol{\varepsilon}}
\nc{\bphi}{\boldsymbol{\varphi}}
\nc{\bpsi}{\boldsymbol{\psi}}
\nc{\bkappa}{\boldsymbol{\kappa}}
\nc{\bom}{\boldsymbol{\omega}}
\nc{\ou}{\boldsymbol{\mathaccent'27u}}
\nc{\oz}{\boldsymbol{\mathaccent'27z}}
\nc{\uu}{\boldsymbol{\breve{u}}}
\nc{\duu}{\dot{\breve{\bu}}}
\nc{\ot}{\mathaccent'27\Theta}
\nc{\utheta}{\breve{\btheta}}
\nc{\otheta}{\boldsymbol{\mathaccent'27\theta}}
\nc{\wt}{\widetilde}

\nc{\IO}{\int_{\Om}} 
\nc{\IQ}[1]{\int_{Q_{#1}}}

\newcommand{\lng}{\raisebox{.52ex}{\rule{12mm}{.1mm}}\kern -4pt}

\nc{\vep}{{\varepsilon}}
\nc{\vr}{{\varrho}}
\nc{\vt}{\vartheta}
\nc{\vfi}{\varphi}

\begin{document}

\title{Solvability of a dynamic rational contact with limited interpenetration for viscoelastic plates}

\author{Ji\v{r}\'{\i} Jaru\v{s}ek\footnote{Institute of Mathematics, Czech Academy of Sciences}}

\date{}

\maketitle

{\small
\noindent {\bf Abstract.} The solvability of the rational contact with limited interpenetration of different kind of viscolastic plates
is proved. The biharmonic plates, von K\'arm\'an plates, Reissner-Mindlin plates and full von K\'arm\'an systems are treated. 
The viscoelasticity can have the classical (``short memory'') form or the form of a certain singular memory. For all models
some convergence of the solutions to the solutions of the Signorini contact is proved provided the thickness of the interpenetration 
tends to zero.\\


}

\section{Introduction and notation}

Despite a great amount of actual and/or possible applications, the theory of contact problems remains still underdeveloped. 
The study of contact problems has been started by A. Signorini \cite{S1}, \cite{S2}. His model describing a contact of a deformable body\
with a rigid foundation respects the impenetrability of Mass. It was extended to dynamic problems by L.~Amerio, G.~Prouse, M.~Schatzman and further 
authors in late seventies and early eighties of the last century. The monograph \cite{ejk} summed up
the development in this field till its publication. The highly nonlinear Signorini model is complex. Therefore a bit later
so called normal compliance approach has been introduced. This approach is nothing else than replacement of the original Signorini contact model
by some kind of its penalization. Although such kind of approximation is a suitable auxiliary tool in
the numerical investigation of contact problems, this approach has  brought no deep results to their theory.
It is usually easy to derive properties of solutions of such approximate problems and the real hard work starts by the limit process to
the original problem.

However the normal compliance approach has drawn the attention to the fact that the complete impenetrability of Mass need not be completely
physically realistic, because from the microscopical point of view no material is flat or smooth enough. Just in the medium advanced microscopes 
the seemingly perfectly flat or smoothly curved surfaces are seen as a huge collections of asperities and small holes or cavities. The asperities may be deformed or may fill the
holes of the counterpart partially or completely. Hence it has some good sense to study models, where some interpenetration between body and the foundation
is allowed to describe macroscopically those phenomena. However, to remain physically realistic, this interpenetration model must include
a certain bound after which the further penetration is not possible. And, as well, it is realistic to assume that such a bound cannot be reached.

These are the premises of the rational contact model which was introduced by \cite{ejs} and \cite{j}, where the solvability
of its static version has been proved. The first dynamic (frictionless) rational contact has been investigated in \cite{js}. It concerns a boundary
contact of a body with a foundation. 

Since 2006 a series of papers about the solvability of dynamic Signorini contact problems for different models of plates \cite{karmve}--\cite{rm} was published. 
The purpose of this paper is to extend these results to the rational contact with limited interpenetration. Unlike \cite{js}
we face here a domain contact.

\section{Abstract formulation of the problem for the clamped or simply supported viscoelastic plate and the scheme of its solution}

Let $\Om \subset \R^2$ be a bounded domain with a sufficiently smooth boundary $\Ga$.
Let $X$ be a Sobolev-type Hilbert space defined on $\Om$, let $Y$ be the space of traces of elements
from $X$ on $\Ga$. Let $A, B: X \to X^*$ be two linear symmetric strongly elliptic operators in the
form $\cD^*a\cD$,  $\cD^*b\cD$, respectively, where $\cD$ is a differential operator and $a$, $b$ are positively definite
matrices  or tensors of time constant but possibly space-dependent elements. Let $I \equiv [0,T]$ be a time
interval.  Here the dual space $X^*$ is defined via the suitable generalization of the $L_2(\Om)$ scalar product.
Let $\cX \equiv L_2(I; X)$. We introduce the bilinear forms $\cA: \{u,v\} \mapsto  \langle a \cD u, \cD v \rangle_Q$,
$\cB:  \{u,v\} \mapsto \langle b \cD u, \cD v \rangle_Q$, where $\langle \cdot, \cdot \rangle_Q$ is the $L_2(Q)$ scalar product
and $Q \equiv I \times \Om$.. Let $S \equiv I \times \Ga$  Let $E(t): X \to X^*$ be anoother operator.  

We shall denote the elements of $v \in X$ or $v: I \to X$ such that $v \in \cX$ as displacements, and their
first time derivatives (denoted by dots) as velocities. Let $\gamma$ be a negative real number. Let $p: \R  \to \bar \R \equiv \R \cup \{+\infty\}$
be a nonincreasing function such that $p(x) =0$ for $x \ge 0$, $pí(x) \in \R$ for $x > \gamma$, and $\lim_{x \searrow \gamma} p(x) = +\infty$,
where $\gamma \in \R$ is a given bound of the interpenetration.
Our problem is to find $u \in \cX$ such that $\dot u \in \cX$ for which the following set of relations holds
\begin{align} \label{cfc} \begin{split} 
  \ddot u &= A \dot u + B u -  E u  + p(u + g) + f \text{ in } X  \text{ on } I,\\
	D(u )&= 0 \in Y, \\
	u(0) &= u_0,\ \dot u(0) = u_1
\end{split} \end{align}
 Here $D$ is a general differential operator of a Dirichlet or somewhat combined type. If $X = H^2(\Om)$, the space of square integrable
functions having the (possibly generalized) first and the second derivatives square integrable as well and $A,\ B$ are differential operators of the fourth order
then $D(u) \equiv \{D_1(u),  D_2(u)\}$, $D_1(u) = u - u_0$  for both cases, $D_2(u)  = \partial_{\tilde n} (u - u_0)$ (the outer {co-)normal derivative)
or $D_2(u) = M(u)$ a Neumann-type operator, which ensures that after the integration
by parts in the space variable in the variational formulation of the problem no additional boundary term occurs. The first couple describes a clamped 
plate while the second one a simply supported plate. Let us mention that $p(u+g)$ stands there for the contact force, where $g \ge 0$ is the gap function, 

We shall define a sequence of auxiliary approximate  problems to \eqref{cfc} by adding the following additional assumption on $p$:
We assume the existence of a sequence $\{\delta_k \} \subset \R_+$ such that $\delta_k \searrow 0$ and for each $k \in \N$ there is a left derivative
$\partial^l p$ in the points $\gamma+\delta_k,\  k \in \N$ such that $\partial^l p(\gamma+\delta_k) \ge \partial^l p(\gamma+\delta_{k+1}), k \in \N$ and
$\lim_{k \to +\infty} \partial^l p(\gamma+\delta_k)  = -\infty$.
Then we define $p_k: y \mapsto \min\{p, p(\gamma+\delta_k)+ \partial^l p(\gamma+\delta_k) (y- \gamma-\delta_k)\} $ for $y\le \gamma + \delta_k$,
$p_k = p$ elsewhere and the auxiliary problem
is defined by replacement of $p$ by $p_k$ in \eqref{cfc}.

Let us denote by $\langle\cdot, \cdot \rangle_\Om$ the duality pairing of $X$ and $X^*$ derived from the $L_2(\Om)$ scalar product  and 
by $\langle\cdot, \cdot \rangle_Q$ the duality pairing of $\cX$ and $\cX^*$ derived from the $L_2(Q)$ scalar product. Let
$\cX_0$ be a subspace of elements of $\cX$ satisfying the appropriate homogeneous Dirichlet boundary condition in \eqref{cfc}, let 
$\cX_1 \equiv \{v \in \cX_0; \dot v \in L_2(Q) \}$. 

Multiplying the first row of \eqref{cfc} by a
test function $v \in \cX_0$ and performing the integration by parts both in space variables and in time we get the variational formulation
of the problem \eqref{cfc}: {\em Find $u \in u_0 + \cX_0$ such that for every $v \in \cX_1$ 
the following equation}
\begin{align}  \label{wcfc} \begin{split}
  &-\langle \dot u, \dot v \rangle_Q +  \langle \cA \dot u, v \rangle_Q +  \langle \cB u, v \rangle_Q +  \langle \cE u, v \rangle _Q - \langle p(u + g), v \rangle_Q +
  \langle \dot u(T, \cdot), v(T, \cdot)\rangle_\Om\\
	&\kern 28ex = \langle f, v \rangle_Q +\langle u_1,v(0,\cdot) \rangle_\Om
\end{split}\end{align}
{\em holds.} For an approximate problem $p$ is replaced by $p_k$ and the integration by parts in time for the acceleration term is omited, hence it
is sufficient to take the test functions from $\cX_0$.

In the sequel we shall assume that the operator $\cE \equiv \{E(t); t \in I\}: \cX \to \cX^*$ is completely continuous, or such that 
$v\mapsto \langle \cE v,v\rangle_Q$ is weakly lower semicontinuous on $\cX_0$,
or such that if a  sequence $v_k \wto v$ in $\cX$ and $\dot v_k \to \dot v$ in $L_2(Q)$, then $\langle \cE v_k, v_k\rangle_Q \to 
\langle \cE v, v \rangle_Q$. 
Moreover, we assume that $\langle \cE v, v\rangle_Q \ge const(u_0, u_1) - c\|v\|_\cX$ for $v \in \cX$ such that $\dot v \in
\cX$ and the initial conditions  in \eqref{cfc} are satisfied. Further, we assume that
\begin{equation} \label{assimp}
  u_0 \in H^2(Q) \text{ such that } u_0 \ge c_0 \text{ on } \bar{Q},\ u_1 \in L_2(\Om) \text{ and } f \in L_2(Q).
\end{equation}
Here $c_0$ is a positive constant.  
	
The proof of the solvability of the auxiliary problem under the assumption \eqref{assimp} does not differ from the proof of a penalized  problem to the
appropriate Signorini contact. It is solved via the Galerkin approximation using just identical arguments, because in this case the auxiliary
contact term represents a completely continuous perturbation of the appropriate problem without contact. By putting $v= (\dot{u}_k - \dot{u}_0) \chi_{Q_t}$
in \eqref{wcfc} with $p_k$, where $\chi_M$ is the characteristic function of a set M (equal 1 on $M$ and vanishing elsewhere), $t \in (0,T]$ and $Q_t \equiv 
[0,t]\times\Om$, we get (after a certain small and obvious calculation) the {\em a priori} estimate of the respective solutions $u_k$ to the approximate
problems with $p_k$ 
\begin{align} \label{ae}
\|\dot u_k\|^2_{L_\infty(I; L_2(\Om))}+ \|u_k\|^2_{L_\infty(I; X)} + \|\dot u_k\|^2_\cX + \|P_k(u_k + g)\|_{L_\infty(I; L_1{\Om})} \le const.,
\end{align}
where $P_k: s \mapsto \int_s^{+\infty} p_k (z)\,dz,\ s \in \R$. Let us take in mind that $L_1(\Om) \subset L_\infty(\Om) ^* \hookrightarrow X^*$,
because for the primal spaces the compact reverse embeddings hold. Since
\begin{align*} 
  \|p_k(u_k + g)\|_{L_1(Q)} \le c_0^{-1} \langle p_k(u_k + g), u_0 - u_k \rangle_Q 
\end{align*}
(observe that $x p_k(x) \le 0,\  x \in \R$), the use of \eqref{wcfc} for $v = u_0 - u_k$ and the estimate \eqref{ae} yields that the sequence 
$\{\|p_k(u_k + g)\|_{L_1(Q)}\}$ is bounded. Then we derive from this and \eqref{cfc} the dual estimate
\begin{equation} \label{de}
  \|\ddot{u}_k\|_{L_1(X^*)} \le const.
\end{equation}
with the constant independent of $k$. With the help of \eqref{de} and the classical Aubin Lemma we get a
certain $u$ and $\vartheta$ such that convergences
\begin{align} \begin{split}
  &u_k \wto^* u \text{ and } \dot{u}_k \wto^* \dot u \text{ in } L_\infty(I; X),\  L_\infty(I; L_2(\Om)), \text{ respectively},
	\dot{u}_k \to \dot u \text{ in } L_2(Q),\\  
	 &\langle \cE u_k, u_k \rangle_Q \to \langle\cE u, u \rangle_Q  \text{ or } \liminf_{k \to \infty}\langle \cE u_k, u_k \rangle_Q \ge \langle\cE u, u \rangle_Q ,
	\text{ and } p_k(u_k+g) \wto \vartheta \text{ in } \cX_1^*
\end{split}
\end{align}
hold for a possible subsequence. Performing the integration by parts in time for the acceleration term and putting $v = u_k - u_0$
in \eqref{wcfc} with $p_k$, using the weak lower semicontinuity of the elliptic operators
and the strong convergence of the others, we get $\langle \vartheta, u \rangle \ge \limsup_{k \to +\infty} \langle p_k(u)_k, u_k \rangle$.
Since $p_k$ are monotone, this yields $\langle \vartheta - p(v + g), u -v \rangle \ge 0$ for every $v \in \cX_0$, hence $[\vartheta,u]$ may be added
to the graph of $p$ such that the extended graph remains monotone.
The maximal monotonicity of $p$ proved in \cite{bz} yields that $\vartheta = p(u + g)$, hence u is a solution of the 
variational equation \eqref{wcfc} and we are done. We have proved

\begin{theorem} 
  Under the above mentioned assumptions to the employed operators and the function $p$ there exists a solution to the problem  \eqref{wcfc}.
\end{theorem} 
 
\noindent {\sf Example 1.} A biharmonic plate. Here $\cD = \triangle,\ a , b$ are positive constants and $\cE =0$. \\

{\noindent \sf Example 2.} A von K\'arm\'an plate without rotation inertia. First we introduce for two functions $u,v$  
\begin{align}
[u, v]  = \partial_{11}\partial_{22}v + \partial_{22}u \partial_{11}v - 2\partial_{12}u \partial_{12}v, 
\end{align}
where here and in the sequel $\partial_i \equiv \partial/\partial_{x_i}$, $i=1,2$, $\partial_t \equiv 
\partial/\partial t$ and $\partial_{ij} \equiv 
\partial_i \partial_j,\ i,j=1,2$. Then we define the bilinear operator $\vf: H^2(\Om)^2 \to \oH^2(\Om)$
by means of the variational equation
\begin{equation} \label{karmop}
  \IO \lap \vf(u,v) \lap \vfi \, dx = \IO [u,v] \vfi \, dx, \ u,v, \vfi \in
  \oH^2(\Om).
\end{equation}
The equation \eqref{karmop} has a unique solution, because $[u,v]
\in L_1(\Om) \emb H^2(\Om)^{*}$. The well-defined operator $\vf$ is
compact and symmetric. Let us recall Lemma 1 from
\cite {ks} due to which $\vf: H^2(\Om)^2 \to W_p^2(\Om)$, for any $p
\in (2,\infty)$, and
\begin{equation} \label{ksest}
\|\vf(u,v)\|_{W_p^2(\Om)}\le c
\|u\|_{H^{2}(\Om)}\|v\|_{W_p^{1}(\Om)}\ \forall\ u,v \in
H^{2}(\Om)^2,
\end{equation}
i.e. $w \mapsto \vf(w,w)$ is completely continuous from $H^{\delta}(Q) \cap \cX$ to $\cX$ for any $\delta>0$.

To avoid the introduction of the Airy stress function, we introduce directly the variational formulation. For it we introduce
\begin{equation} \label{kbf}
  A_0 : \{u,y\} \mapsto b_0 \big(\partial_{\ell\ell} u \partial_{\ell\ell} y +
  \nu(\partial_{11}u \partial_{22}y + \partial_{22}u \partial_{11} y)
  + 2(1-\nu)\partial_{12}u \partial_{12}y \big), b_0= const>0,
\end{equation}
where $\nu \in (-1/2, 1)$ is a material constant (the Poisson ratio) and the standard summation convention for the repeating index $\ell$
is applied.  Then we define $\langle \cA \dot u, v\rangle_Q$
as $e_1\int_Q A_0(\dot u, v) dx\, dt$,   $\langle \cB u, v\rangle_Q $ as  $e_0 \int_Q A_0( u, v) dx\, dt$,
$\cE: u \mapsto b([u, e_1\partial_t \lap\vf(u,u) + e_0 \lap \vf(u,u)])$, where $e_1$, $e_0$ are other material constants (the Young moduli)
which are positive. With such defined mappings the variational formulation of the problem has exactly the form of \eqref{wcfc}.
It is easy to derive that 
\begin{equation} \label{cc}
 \langle \cE u_k, u_k  \rangle_Q = \int_Q  \big( e_1/2\, \partial_t  (\lap\vf(u_k,u_k) )^2 + e_0 (\lap\vf(u_k, u_k) )^2 \big) dx\,dt
\end{equation}
(cf. \cite{karmve}) hence it satisfies the corresponding requirements and the quadratic forms generated by such defined $\cA, \cB,
\langle \cE\cdot, \cdot \rangle_Q$
are weakly lower semicontinuous and we are done.
We remark that $M(u) = b( e_1 m(\dot u) + e_0 m(u))$, where $m(u) =
\triangle u + (1-\nu )\big(2 n_1 n_2 \partial_{12}u - n_1^2\partial_{22}u - n_2^2\partial_{11}u \big)$. 
\\
 
\noindent {\sf Example 3,} A simply supported von K\'arm\'an plate with the rotation inertia. Here the original structure \eqref{cfc} is enriched by
the additional term $Gu = g_0 \triangle \ddot u$ to the right hand side of the first row of \eqref{cfc}. If $g_0$ is just a positive constant, 
then this term contributes (after the obvious integration by parts) to the extension of the a priori estimate \eqref{ae} by the term
$\|\nabla \dot{u}_k \|^2_{L_2(I;L_2(\Om))}$. The dual estimate $\|g\lap \ddot u_k - \ddot u_k\|_{L_2(I,.X^*)}\le const$ is here $k$-dependent.
After integration by parts this gives $\sup_{v \in L_2(I;X), \|v\| \le 1} \langle \ddot u, g\lap v - v\rangle_Q \le const$ . The operator $g_0 \lap - I$,
where $I$ is the identity, is an isometry between the space $X = H^2(\Om) \cap \oH^1(\Om)$ and $L_2(Q)$, hence the dual estimate yields
$\ddot{u}_k \in L_2(Q)$. In  the further treatment an additional lower semicontinuous term of this form occurs which does
not change the treatment of the limit process from the approximate to the original problem. In fact from  the $k$-independent $L_1(Q)$ 
estimate of the approximate contact term we get (using again the properties of the operator $(g_0 \lap - I$) the $k$-independent dual estimate
$\|\ddot{u}_k \|_{L_1(I;L_2(\Om))} \le const$. The Aubin Lemma again yields the crucial strong convergence
$\dot{u}_k$ to $\dot u$ in $L_2(Q)$. Hence we are in the same situation as above, $u_k \mapsto \langle p_k(u_k + g), u_k \rangle_Q$ is again upper semicontinuous 
and via the maximal monotonicity argument we are done. 

\section{Von K\'arm\'an model with a singular memory}

Let us introduce the kernel $K$ of the singular memory term which is assumed to be integrable over $\R_{+}$ and to have the form
\begin{align} \label{memintro}  \begin{split}
  K&: t \mapsto t^{-2\alpha}q(t) + r(t),\ t \in \R_{+}
  \equiv ( 0, +\infty)
  \hbox{ with } \alpha \in \left(0, \tfrac{1}{2}\right),\\
  K&: t \mapsto 0,\ t \le 0.
\end{split}  \end{align}
Both $q$ and $r$ belong to $C^{1}(\R_+)$; they are non-negative and non-increasing functions. Moreover, we assume that
$q(t) > 0$ for $t$ on an nonempty interval $[0,t_0]$. Let $d_m: v \mapsto \int_{0}^{t} K(t - s)\big( v(t,\cdot) -  v(s,\cdot)\big)\, ds$
for a function $v$ on $Q$. Let us remark that it holds
\begin{align}   \begin{split} 
   \langle d_m v, \dot v \rangle_Q &= \int_Q \int_s^T \frac 1 2  \big(\partial_t \big(K(t-s)(v(t) - v(s))^2\big) - (v(t)-v(s))^2 \partial_t K(t-s) \big) dx\,dt\,ds \\
  &= \int_Q \frac 1 2 K(T-s) (v(T) - v(s))^2 dx\,ds  - \int_Q \int_0^t \frac 1 2 K'(t-s)(v(t)-v(s))^2 dx\,dt\,ds
\end{split} \end{align} 
and the second term in this formula leads to the fractional time--derivative norm of $v$. We recall that such a norm used in the sequel is
for a Banach space $X$ defined  as follows:
\begin{equation*}
\| v \|_{H^\alpha(I;X)}^2\equiv\int_I \|v\|_X^2dt+\int_I\int_I
\frac{\|v(t)-v(s)\|_X^2}{|t-s|^{1+2\alpha}}\,ds\,dt.
\end{equation*}

We solve the problem
\begin{align} \label{kmsyst}  \begin{split}
\ddot{u} - e_1d_m Au - e_0 A u + \cE_0 u = f + p(u + g) \text{ on } Q, \\
u = 0, \quad M(u) = 0   \text{ on } S,
u = u_0,  \dot u  = u_1 \text{ on } \Om,
\end{split}  \end{align}
Here $A$ is the differential operator  leading to the operator defined in \eqref{kbf}, i .e.
\begin{equation}
  b_0 \big( \partial_{\ell\ell} \partial_{\ell\ell}  +   \nu(\partial_{11} \partial_{22} + \partial_{22} \partial_{11} )
  + 2(1-\nu)\partial_{12} \partial_{12} \big) 
\end{equation}
and
\begin{align*} \begin{split}
\cE_0:  u  & \mapsto [u,e_1 d_m 
\lap \vf(u,u) + e_0 \lap \vf(u,u)] , \\
M(u) &= a d_m m(u) + b m(u)], \text{ where } m(u) = \triangle u + (1-\nu )\big(2 n_1 n_2 \partial_{1,2}u -
n_1^2\partial_{2,2}u - n_2^2\partial_{1,1}u \big) \\
&\text{for a simply supported plate.}\\
M(u) &= \partial u/ \partial n \text{ for the clamped plate.}  
\end{split} \end{align*}

To be able to handle the singular memory term it is necessary to assume its smallness as follows
\begin{equation} \label{smallmem}
  \int_{0}^{+\infty} K(s) \, ds < e_0/2e_1
\end{equation}
which ensures that the quadratic form 
\begin{equation}
Z: V \mapsto \int_Q (e_1 d_m V + e_0 V) V dx\,dt,\ V \in L_2(Q)
\end{equation}
is strongly monotone.

We introduce the variational formulation of the problem. Let $X = H^2(\Om) \cap \oH^1(\Om)$ 
for the simply supported plate and $X = \oH^2(\Om)$ for the clamped plate. 
The formulation reads: {\em Find $u \in L_2(I; X) \cap H^1(I; H^1(\Om))$
such that for every $v \in L_2(I; X_0) \cap H^1(I; H^1(\Om))$ the equation}
\begin{align} \begin{split}  \label{vfksm}
   &\IQ{} \big(  (e_1 d_m A_0(u,y) + e_0 A_0 (u,y)   - \dot u \dot y + \cE_0 u y - p(u + g)y - f y \big) \,dx \, dt \\
      &\kern 2em + \IO \big(-(\dot{u} y)(T,\cdot)  +  u_1(y(0,\cdot)  \big) \, dx = 0 
\end{split}  \end{align}
{\em holds with $A_0$. from \eqref{kbf}.}  We shall solve this problem assuming that \eqref{assimp} holds.

We formulate the approximate problem again by replacing $p$ by $p_k$, but unlike \eqref{vfksm} no integration by parts in 
time for the acceleration term is applied, hence the test function may be taken just from $\cX_0$. It is solved again by 
the standard Galerkin procedure, for details cf. \cite{karmsg}.
To get the $k$--independent {\em a priori} estimate for their solution,
$y = \dot u_k -\dot u_0$ must be taken. After some calculation we finally obtain
\begin{align} \begin{split} \label{gae}
& \| u_k\|^2_{H^{\alpha}(I; H^2(\Om))} +
 \|\dot{u}_k\|^2_{L_{\infty}(I;L_21(\Om))} +
 \|u_k\|^2_{L_{\infty}(I;H^{2}(\Om))} +
 \| \vf(u_k,u_k)\|^2_{H^{\alpha}
(I;H^2(\Om))}\\ &+ \|P_k(u + g)\|_{L_\infty(I;L_1(\Om))}
 \le c \equiv c(f,u_0,u_1),
\end{split}\end{align}

Since the singular--memory terms are remarkably weaker than the corresponding viscoelastic ones,
the Aubin Lemma gives not a sufficient reasoning to prove the required strong convergence  of velocities
in the limit procedure for $k \to \infty$. In the sequel we shall use the following theorems:

\begin{theorem}[{\rm Embedding theorem}]  \label{embth}
 Let $M \subset \R^{N}$ be a bounded domain with a Lipschitz
  boundary. Let $p,q \in (1,\infty)$, $\gamma \in [0,1]$
  and $\alpha
  \in (\gamma, 1]$ be numbers such that the inequality
  \begin{equation} \label{imbin}
   \frac{1}{\alpha}\left( \frac{N}{p} - \frac{N}{q}
   + \gamma \right) \le 1,
  \end{equation}
holds. Then the Sobolev--Slobodetskii space $W^{\alpha}_{p}(M)$ is
continuously embedded into $W^{\gamma}_{q}(M)$. If inequality
\eqref{imbin} is strict, then the embedding is compact for any
real $q \ge 1$. For $q= \infty$ this is true under the convention
$1/q =0$.
\end{theorem}

\begin{cor} \label{emboch}
Let $M$ and $I$ be as above. Let $p_i,q_i$ belong to
$(1,+\infty)$, $\alpha_i$ belong to $(0,1]$ and $\gamma_i$ to
$[0,\alpha_i)$, $i=1,2$. Assume that \eqref{imbin} holds with
$i=1$ and $N$ replaced by $1$  and that it simultaneously holds
for $i=2$. Then $W^{\alpha_1}_{p_1}\!\left(I;
W^{\alpha_2}_{p_2}(M)\right)$ can be imbedded into
$W^{\gamma_1}_{q_1}\!\left(I; W^{\gamma_2}_{q_2}(M)\right)$. If both
inequalities are strict, the imbedding is compact. The last
assertion still holds if $q_i$ is infinite, provided we use the
convention $1/q_i =0$, $i=1,2$.
\end{cor}

\begin{theorem}[{\rm Interpolation theorem}] \label{interth}
Let $M$ be as above, let $k_{1}$, $k_{2}$ belong to $[0,1]$, let
$p_{1}$, $p_{2}$ belong to $(1,+\infty)$ and $\Theta_\lambda$ to
$[0,1]$. Then there exists a constant $c$ such that for all $u \in
W^{k_1}_{p_1}(M) \cap W^{k_2}_{p_2}(M)$ the following estimate
holds
$$
\|u\|_{W_{p}^{k}(M)} \le c
\|u\|_{W_{p_1}^{k_1}(M)}^{\Theta_\lambda}
\|u\|_{W_{p_2}^{k_2}(M)}^{1 -\Theta_\lambda}  $$ with $k =
\Theta_\lambda k_1 + (1-\Theta_\lambda) k_2$ and $\dfrac{1}{p} =
\dfrac{\Theta_\lambda}{p_1} + \dfrac{1-\Theta_\lambda}{p_2}$. The
assertion remains true if $k_1 = k_2=0$ and $p_1$, $p_2$ belong to
$[1, +\infty]$.
\end{theorem}

\begin{cor}[{\rm Generalization}] \label{interboch}
Let $M$, $k_{1}$, $k_{2}$, $p_{1}$, $p_{2}$ be as above. Let $I$
be a bounded interval in $\R$, let $\kappa_{1}$, $\kappa_2$ belong
to $[0,1]$, let $q_{1}$, $q_{2}$ belong to $(1, +\infty)$ and
$\Theta_\lambda$ to $[0,1]$. Then there exists a constant $c$ such
that for all $u \in W^{\kappa_1}_{q_1}\!\!\left(I;
W^{k_1}_{p_1}(M)\right) \cap W^{\kappa_2}_{q_2}\!\!\left(I;
W^{k_2}_{p_2}(M)\right)$ it holds
$$\|u\|_{W^{\kappa}_{q}\!\left(\!I;
 W_{p}^{k}(M)\!\right)} \le c
\|u\|_{W^{\kappa_1}_{q_1}\!\left(\!I;
W_{p_1}^{k_1}(M)\!\right)}^{\Theta_\lambda}
\|u\|_{W^{\kappa_2}_{q_2}\left(\!I; W_{p_2}^{k_2}(M)\!\right)}^{1
-\Theta_\lambda}, $$ where $k = \Theta_\lambda k_1 +
(1-\Theta_\lambda) k_2$, $\kappa = \Theta_\lambda \kappa_1 +
(1-\Theta_\lambda) \kappa_2$, $\dfrac{1}{q} =
\dfrac{\Theta_\lambda}{q_1} + \dfrac{1-\Theta_\lambda}{q_2}$ and
$\dfrac{1}{p} = \dfrac{\Theta_\lambda}{p_1} + \dfrac{1 -
\Theta_\lambda}{p_2}$. If $\kappa_1 = \kappa_2=0$ and $q_{1}$,
$q_2$ belong to $[1,+\infty]$, the assertion still holds.
\end{cor}

The proofs of the above mentioned facts follows from Chapter 2 of the monograph
\cite{ejk}.\\

The assumed  smallness of the memory term yields again the uniform estimate of
$\{ \|p_k(u+g)\|_{L_1(Q)} \}$ which  leads to the dual estimate 
$\|\ddot u_k\|_{L_1(I; X^*)} \le const$. Hence $\{\dot{u}_k\}$ is bounded in
$W^{1-\vep_1}_{1+\vep_2}(I;H^{-2-\vep_3}(\Omega ))$ for any $\vep_2>0,
\vep_3 > 0$ and for $\vep_1 \equiv \vep_1(\vep_2) \searrow 0$ if $\vep_2
\searrow 0$. Interpolating this space with the space $L_{q}(I;
L_2(\Omega ))$ for $q  = 1+1/\vep_2$ we get that
\begin{equation} \label{interest}
  \|\dot{u}_k\|_{H^{1/2}(I;H^{-1-\theta}(\Omega ))} \le C, \text{ i. e. }
	\|u_k\|_{H^{3/2}(I; H^{-1-\theta}(\Omega ))} \le C
  \text{ with } 0< \theta \text{ arbitrarily small.}
\end{equation}
Interpolating the result in \eqref{interest} with the fact that
$\{u_k\}$ is bounded in $H^{\alpha}(I; H^{2}(\Omega))$, we get that
$\{\dot u_k\}$ is bounded in $H^{\theta_1}(I; L_2(\Omega ))$ 
for $\theta_1, \in (0, \alpha/3)$. Interpolation of this space with the time-fractional
derivative space from \eqref{gae} gives the space $L_2(I; H^{\delta_2})$ with
$\delta_2 \in (0, 2\alpha/(3-2\alpha))$, hence $\{\dot u_k\}$ is bounded in the 
anizotropic space $H^{\theta_1, \theta_2}(Q)$. This space is compactly
imbedded into $L_2(Q)$ which ensures that $\langle \dot{u}_k, \dot{u}_k  \rangle_Q$ tends strongly
to the limit $\langle \dot u, \dot u \rangle_Q$ even
for the weak convergence of $u_k$ in the employed spaces. Similarly to \eqref{cc}
we can derive that
$$
   \langle \cE_0 u_k, u_k  \rangle_Q = \int_Q ( e_1 \lap d_m\vf(u_k, u_k) \lap\vf(u_k,u_k) \  + e_0 (\lap\vf (u_k,u_k))^2  ) dx\,dt.
$$
The compactness of $\vf$ based on \eqref{ksest} and the fractional time-derivative norm in \eqref{gae} yield the needed strong convergence of this
term. Hence we are able for the limit procedure $k \to +\infty$ to prove again the upper semicontinuity of $\langle p_k(u_k), u_k \rangle_Q$  and
with the maximal monotonicity argument to prove $p_k(u_k + g) \wto p(u + g)$. Thus $u$ is a solution of \eqref{vfksm} and with 
the additional assumption \eqref{smallmem} the existence theorem is proved also for this problem.

\section{The problem for more complex viscoelastic plate models}

In this section we shall treat the Reissner-Mindlin plate model as well as full von K\'arm\'an system. The plates are again 
in contact with the limited interpenetration with a foundation. 

\subsection{Contact of Reissner-Mindlin plates}  

In this 2nd order model besides the vertical deflection $u$ there is the 2D--vector $\bphi$ of angles
of rotations of the cross sections of the plate. We denote by $\bbS$ the set of symmetric $2\times 2$
tensors with the product ${\boldsymbol \kappa}\odot {\boldsymbol \lambda}=\kappa_{ij}\lambda_{ij}$, where the Einstein
summation convention (summing over repeated indices) is employed. Moreover for $\omega \equiv 
\{\omega_{ij},\ i,j=1,2 \} \in \bbS$ we denote $\Div \omega \equiv (\partial_i \omega_{1i}, \partial_i \omega_{2i} )$
and   $\tr \omega = \omega_{11}  + \omega_{22}$.

With the notation
\begin{align}  \label{rmnot}  \begin{split}
  J(u,\bphi) &= e_1(\nabla \dot u+\dot\bphi)+e_0(\nabla u + \bphi), \\
	 {\mathcal C}_i(\omega) & =\frac{\tilde c}{(1-\nu_i^2)}
\big(\nu_i(\mathrm{tr}\,\omega)I_\mathbb{S}+(1-\nu_i)\omega\big),\ \omega\in \mathbb S,\  i=0,\,1,
\end{split} \end{align}
where  $I_\mathbb{S}$ is the unit matrix in $\mathbb{S}$, 
$\tilde c, e_0, e_1$ are given positive constants, and the
Poisson ratio $\nu_i \in (-1, 1/2)$,  $i = 0,1$ {\em the classical formulation of the viscoelastic (``short memory'') problem} is as follows:
We look for $ (u, \bphi)$ such that the system 
\begin{align}  \left. \begin{array}{rl}  
\ddot{u} - \mdiv\ J(u, \bphi) &= f + p(u + g),\\[.5ex]
\ddot{\bphi}-\Div\big({\mathcal C}_1(\bvep_0(\dot\bphi))+{\mathcal C}_0(\bvep_0(\bphi))\big)+J(u, \bphi)  &=\bM
\end{array} \right\} &\text{ on } Q,\label{rm}
\end{align}
the boundary value conditions
\begin{align} \label{bcrm}
\left. \begin{array}{rl}
u = u_0,\ \bph = \boldsymbol{0} & \text{ for a clamped plate,} \\[1ex]
u = u_0,\ \big({\mathcal C}_1(\bvep_0(\dot\bphi))+{\mathcal C}_0\bvep_0(\bphi)\big) \cdot \bn =\boldsymbol{0}
& \text{ for a simply supported one}
\end{array}\right\}\text{ on } S,
\end{align}
 and  the initial conditions
\begin{align} \label{initrm} \left. 
\begin{array}{l}
  u(0, \cdot)= u_0 ,\quad   \dot{u}(0,\cdot)=u_1, \\[1ex] 
 \bphi(0,\cdot)= \bphi^{(0)},\ \dot{\bphi}(0, \cdot) = \bphi^{(1)}
\end{array}  \right\} & \text{ on } \Om 
\end{align}
are satisfied. Here $\bvep_0$ is the standard 2D linearized strain tensor and 
$\bn$ is the unit outer normal vector. We assume that the function $p$
satisfies all the assumptions listed at the beginning of Section 2,  we assume that \eqref{assimp} still holds,
in particular  the positive function $u_0$ is again bounded away from 0. Moreover, we assume that
$\bphi^{(1)} \in \bL_2(\Om)$, $\bphi^{(0)} \in \bH^1(\Om)$ and $\bM \in \bL_2(Q)$.

The variational formulation of the problem based on appropriate integrations by parts  
has the following form: {\em Look for $\{u,\bphi\} \in (u_0 + L_2(I; \oH^1(\Om))\times
\bX(Q)$ such that $\dot u\in L_2(I; H^1(\Om)),\ \dot \bphi\in
L_2(I;\bX(\Om)), \ \ddot\bphi\in \bL^2(Q)$, the first condition in the first row and the second row of the initial conditions \eqref{initrm}
are satisfied  and the system}
	\begin{align}\begin{split}\label{vfrm}
   & \IQ{ }  \big( J(u,\bph) \cdot \nabla y - \dot u  \dot y - p(u + g) y \big)\,dx \,dt =  \IO{ }\big( u_1 \,  y(0,\cdot) -\dot u(T,\cdot) \, y(T,\cdot)\big)\,dx
	 + \IQ{ } f  y \,dx
\,dt,\\\
     & \IQ{ }\big(\ddot\bphi\cdot \bpsi + \big({\mathcal
C}_1(\bvep_0(\dot\bphi))+{\mathcal
C}_0(\bvep_0(\bphi))\big)\odot\bvep_0(\bpsi)+ J(u, \bph) \cdot \bpsi\big)\,dx \,dt =\IQ{ }\bM\cdot\bpsi \,dx \,dt
\end{split}
\end{align}
{\em holds for any $\{y,\bpsi\} \in   \oH^1(Q) \times L_2(I,\bX (\Om)) $.} Here $\bX$ stands for $\obH^1$, $\bH^1$
for clamped and simply supported plates, respectively.

As in previous cases we introduce the approximate problems by replacing the original function $p$ by the
approximate function $p_k$ and the integration by parts in time for the acceleration term in the first row of
\eqref{vfrm} is omitted. Hence it has the form
\begin{align}  \label{vfrma}
   & \IQ{ }  \big( J(u_k,\bph_k)  \cdot \nabla y + \ddot{u}_k  y \big)\,dx \,dt =  	\IQ{ } (f +  p_k(u_k + g) ) y \,dx
\,dt.
\end{align}
We put $\{y,\bpsi\} = \{ \dot{u}_k -\dot{u}_0, \dot{\bph}_k \}$ as the test function of the approximate system 
and integrate on the interval $[0,s],\ s\le T$.
Adding both lines of \eqref{vfrm} and using the standard integration by parts we get
\begin{align} \begin{split} \label{deraerm}
   &\IQ{s} \left( \frac 12 \partial_t
 \big( \dot{u}_k^2
     +  d_0|\nabla u_k+\bphi_k|^2+ \vert\dot\bphi_k\vert^2
     +\mathcal{C}_0(\bvep_0(\bphi_k))\odot\bvep_0(\bphi_k)
 + P_k(u_k + g) \big) \right. \\
  & +d_1|\nabla\dot u_k+\dot \bphi_k|^2
     +\mathcal{C}_1\big(\bvep_0(\dot \bphi_k)\big)\odot\bvep_0(\dot\bphi_k) \Big)\, dx\, dt
= \IQ{s}\big(f\dot u_k+\bM\dot\bphi_k\big)\, dx \,dt\\
  &\kern 12 em + \IQ{s} R(\dot u_0) \, dx \,dt,
\end{split}
\end{align}
where in $R(u_0)$ we sum up all the terms containing $\dot u_0$, or its derivatives.
From the positive definiteness of the tensors $\mathcal{C}_i$ and the last identity we derive after
some calculation the {\em a priori} estimate
\begin{align}\begin{split}
   &  \|\dot{u}_k\|^2_{L_\infty(I;L_2(\Om))}+\|\dot{\bphi}_k\|_{L_\infty(I;\bL_2(\Om))} +
	\|\dot{u}_k\|^2_{L_2(I;H^1(\Om))}+\|\dot{\bphi}_k\|^2_{L_2(I;\bH^1(\Om))} +
    \|u_k\|^2_{C(\bar I;H^{1}(\Om))} \\
    &  +\|{\bphi}_k\|^2_{C(\bar I;\bH^1(\Om))} +  \|P_k(u_k + g)\|_{L_\infty( I;L_1(\Om))} 
     \le c \equiv c\big( f,\bM,u_0,u_1,\bphi^{(0)},\bphi^{(1)} \big).\end{split}\label{aerm}
\end{align}
Observe that this estimate is $k$-independent. 

We continue with the estimates of the acceleration terms. After using $\{\ddot{u}_k -\ddot{u}_0, \ddot\bph_k\}$ 
as the test function we obtain
\begin{align}
&\label{ddotbph}\|\ddot{\bphi}_k\|^2_{\bL_2(Q)}\le c,\\
&\label{ddotu}
\|\ddot{u}_k\|^2_{L_2(Q)}\le c_k,\ k\in \N.
\end{align}
From \eqref{aerm}  it is easy to see that \eqref{ddotbph} is again $k$--independent. However, \eqref{ddotu} 
depends on $k$ and for the limit process $k \to \infty$ it must be replaced by the dual estimate
of $\ddot{u}_k$ based on the estimate of the contact term.

These approximate problems are solved by means of the Galerkin approximation. Since they do not structurally
differ from the penalized problems for the Signorini contact (in both described cases the approximate contact term
represents a compact perturbation of the non--contact problems) and we are focused here on the difference
between the rational contact with  limited interpenetration and the Signorini contact, we omit details of this well--known 
process here and postpone the readers to \cite{rm} for them.

To derive the crucial dual estimate of $\ddot{u}_k$ we can use the general abstract approach of Section~2. 
However, the space $H^1(\Om)$ is not imbedded into $L_\infty(\Om)$, hence we must use $X = H^1(\Om)
\cap L_\infty(\Om)$ here. The resulting estimate \eqref{de} yields the required strong convergence of 
$\dot{u}_k$ in $L_2(Q)$ in the process $k \to \infty$ via the Aubin Lemma. We put $\{y,\bpsi\} = \{u_k - u_0, \bph_k\}$ 
in \eqref{vfrm} and add both equations. We get
\begin{align}  \label{derrm} \begin{split}
   &\IQ{T} \big( -\dot{u}_k^2      +  e_0|\nabla u_k+\bphi_k|^2+ \vert\dot\bphi_k\vert^2
     +\mathcal{C}_0(\bvep_0(\bphi_k))\odot\bvep_0(\bphi_k)
  + p_k(u_k + g)(u_k - u_0) \big)  \\
  & + \partial_t  \big(  e                                                                                                                                                                                                                                                               _1|\nabla u_k+ \bphi_k|^2
  +\mathcal{C}_1\big(\bvep_0(\bphi_k)\big)\odot \bvep_0(\bphi_k) \big)  \emph{}\, dx\, dt
 =   \IQ{T}\big(f\dot u_k+\bM\dot\bphi_k\big)  \, dx \,dt\\
  &\kern 12 em + \IQ{T} R_1(u_0, u_1) \, dx \,dt,
\end{split} \end{align}
where $R_1$ contains all remaining terms. Obviously they contain $u_0$ or $u_1$ or their derivatives. This identity shows again 
that it belongs to the abstract structure described in Section~2. Besides weakly lower semicontinuous
elliptic terms and weakly continuous terms as  $\dot{u}_k^2$ and $R_1$ the only remaining term,
the contact one, must be upper semicontinuous and we can use again the maximal monotonicity argument 
for it to prove $p_k(u_k+g) \wto p(u+g)$. Hence the limit $u$ of the sequence $\{u_k\}$ satisfies \eqref{vfrm} and we are done.\\

 In the classical formulation of the Reissner--Mindlin plate with a singular memory we replace all the ``short memory''
terms in $J$ and ${\mathcal C}_1$ (i.e. the terms containing the time derivatives) by the corresponding singular
memory terms (the $d_m$ versions of the elastic terms), where we use again the kernel $K$ defined in \eqref{memintro}.
Hence $J(u, \bph) \equiv e_0(\nabla u + \bph) + e_1d_m(\nabla u + \bph)$. 
With this modification the structure of \eqref{rm}, \eqref{bcrm}, and \eqref{initrm} remains preserved. We assume
again the sufficient smallness of the memory. To get it exactly in the form \eqref{smallmem} we assume $\nu_1 = \nu_0$. 

We present explicitely its variational formulation which reads:  {\em Look for $\{u,\bphi\} \in (u_0 + \oH^1(Q)) \times L_2(I; \bX(\Om))$ 
such that $\ddot\bphi\in \bL^2(Q)$, the first condition in the first row and the second row of \eqref{initrm}
are satisfied  and the system}
	\begin{align}       \label{vfrmem}
		& \IQ{ }  \big( J(u,\bph) \cdot \nabla y - \dot u  \dot y \big)\,dx \,dt =  \IO{ }\big( u_1 \,  y(0,\cdot) -\dot u(T,\cdot) \, y(T,\cdot)\big)\,dx
	 + \IQ{ } (f + p(u + g) ) y \,dx
\,dt, \nonumber    \\
     & \IQ{ }\big(\ddot\bphi\cdot \bpsi + ({\mathcal C}_1(d_m\bphi) + \cC_0(\bphi))  \ \odot\bvep_0(\bpsi)+ 
		J(u, \bph) \cdot \bpsi\big)\,dx \,dt =\IQ{ }\bM\cdot\bpsi \,dx \,dt
\end{align}
{\em holds for any $\{y,\bpsi\} \in  L_2(I; H^1(\Om))  \times L_2(I,\bX (\Om)) $.} Here again $\bX$ stands for $\obH^1$, $\bH^1$
for clamped and simply supported plates, respectively.

We formulate again the approximate problems by replacing the function $p$ by $p_k$ and by omitting the 
integration by parts at the acceleration  term. We solve this problem via the Galerkin method as usually. 
We again omit here the details postponing the readers to the paper \cite{rm}. Since it is
not clear at the beginning whether the velocity $\dot{u}_k$  possesses the requred  qualites of the test
function, the apriori estimates there have been derived for the finite--dimensional space approximations
and then the limit process to the original infinite--dimensional space has been performed. However,
the result is the same as if we put formally  $\{\dot{u}_k, \dot{\bph}_k\}$ as the test function. 

Summing up 
both equations  and limiting the integration to the cylinder $Q_s$ for  $s\le T$ we obtain using the
 properties  of the kernel function $K$ the identity
\begin{align} \begin{split} \label{mempid1}
   \IQ{s} \Big(& \frac 12 \partial_t
 \big( \dot{u}_k^2
     +  d_0 |\nabla u_k+\bphi_k|^2+ \vert\dot\bphi_k\vert^2
     + \mathcal{C} (\bvep_0(\bphi_k))\odot\bvep_0(\bphi_k)
 + P_k(u_k + g)   \big)
 \\&
+\frac d2K(s-t)|\nabla (u_k(s)- u_k(t))+\bphi_k(s)-\bphi_k(t)|^2
 \\
&+\frac b2K(s-t)\mathcal{C}(\bvep_0(\bphi_k(s)-\bphi_k(t))\odot\bvep_0(\bphi_k(s)-\bphi_k(t))\Big)\, dx\, dt\\
 &-\frac d2\IQ{s} \int_0^t K_t'(t-\tau)|\nabla (u_k(t)- u_k(\tau))+\bphi_k(t)-\bphi_k(\tau)|^2
\, d\tau\, dx\, dt\\
&-\frac b2\IQ{s} \int_0^t K_t'(t-\tau)
\mathcal{C}(\bvep_0(\bphi_k(t)-\bphi_k(\tau))\odot\bvep_0(\bphi_k(t)-\bphi_k(\tau))
\, d\tau\, dx\, dt
\\ &= \IQ{s}\big(f\dot u_k+\bM\dot\bphi_k\big)\, dx \,dt.
\end{split}
\end{align}
By virtue of \eqref{memintro}, \eqref{smallmem} the identity
\eqref{mempid1} leads to the {\em a priori} estimates independent of
of $k\in \N$:
\begin{align} \begin{split} \label{gaem}
& \|\dot{u}_k\|^2_{L_{\infty}(I;L_2(\Om))} +\|\dot{\bphi}_k\|^2_{L_{\infty}(
I;\bL_2(\Om))}+\| u_k\|^2_{H^{\alpha}(I; H^1(\Om))} +\|\bphi_k
\|^2_{H^{\alpha}(I; \bH^1(\Om))} + \|u_k\|^2_{L_\infty(I;H^{1}(\Om))} \\
& + \|\bphi_k\|^2_{L_\infty(I;\bH^{1}(\Om))}+\|P_k(u_k + g)\|_ {L_\infty(I;L_1(\Om))}  \le c
\equiv c \big(f,\bM,u^{(0)},u^{(1)},\bphi^{(0)},\bphi^{(1)}\big).
\end{split}\end{align}
 
The estimate of the accelerations is a straightforward consequence
of the {\em a priori} estimate \eqref{gaem} and the approximate system to
\eqref{vfrmem} and has the form:
\begin{align}
&\label{ddotphim}\|\ddot{\bphi}_k\|^2_{\bL_2(I;(H^1(\Om))^*)}\le c,\\
&\label{ddotum}
\|\ddot{u}_k\|_{L_2(I;(H^1(\Om))^*)}^2\le c_k
\end{align}
(the first of them is again $k$--independent).

For the limit process $k \to \infty$ we can  get  the dual estimate 
$\|\ddot{u}_k\|_{L_1(I; H^{-1-\tilde\vep}(\Om))} \le const.$ via the $L_1$
estimate of the approximate contact term (cf. \eqref{de})  if we
employ the dual embedding $H^{1 + \tilde\vep}(\Om) \emb L_{\infty}(\Om)$ for
any $\tilde\vep >0$.  Then the sequence $\{\dot u_k\}$ is  bounded in
$W_{1+\vep_2}^{1-\vep_1}(I;H^{-1- \tilde\vep}(\Om))$ for any $\vep_2 >0$ 
and $\vep_1\equiv \vep_1(\vep_2)\searrow 0$ if
 $\vep_2\searrow 0$. Simultaneously it is bounded in
 $L_q(I;L_2(\Om))$ for any $q \ge 2.$ After interpolating the spaces
$W_{1+\vep_2}^{1-\vep_1}(I;H^{-1 - \vep_0}(\Om))$ and $L_q(I;L_2(\Om))$
for $q \ge 1+1/\vep_2$  we have
\begin{equation*}
\| \dot u_k \|_{H^{1/2}(I;H^{-1/2 - \vep_3}(\Om))}\le const., \text{ i. e. } 
\Vert u_k \Vert_{H^{3/2}(I;H^{-1/2-\vep_3}(\Om))}\le const., \ k \in \N,
\label{muestim}
\end{equation*}
where $\vep_3>0$ is arbitrarily small..
Interpolating this result with the fact that $\{u_k\}$ is bounded in
$H^{\alpha}(I; H^1(\Om))$ for the given $\alpha \in (0,\frac 1 2)$, we obtain that $\{u_k\}$ is
bounded in the space $H^{1+\delta_1}(I; L_2(\Om))$ for any $\delta_1 \in (0, \alpha/3)$.
Interpolating this result  with the same space we get the boundedness of $\{u_k\}$ in
$H^1(I; H^{\delta_2}(\Om))$ for any $\delta_2 \in \big(0, \alpha/(3-2\alpha)\big)$. The intersection of both 
resulting spaces is obviously compactly imbedded into $H^1(I; L_2(\Om))$, hence the strong convergence of the velocities
is proved.  As earlier, the resulting upper semicontinuity of the contact term and the maximal monotonicity argument for $p$
leads to the fact that $p_k(u_k+g) \wto p(u + g)$ and the existence of a solution to the system \eqref{vfrmem} is thus ensured.

\subsection{Contact of viscoelastic plates described by full von K\'arm\'an system}

This model of plates describes the vertical deflection $u$ as  well as the horizontal ones denoted by
$\bu \equiv \{u_1 , u_2\}$. We assume that the potential contact is both with the foundation of the plate and
on the boundary $\Ga$. We preserve the notation of $\cC_i $ from \eqref{rmnot}, but the physical meaning of
some terms may differ here from the previous parts.
Denoting 
\begin{equation}\label{stress}
\mathfrak{C}_0  = e_0 \rs{C}_0 \big(\varepsilon(\bu)+\Psi(\nabla u) \big),\
 \mathfrak{C}_1 = e_1 \rs{C}_1 \big(\varepsilon(\dot \bu)+ \partial_t \Psi(\nabla u) \big),\
 \Psi(\ba)=\frac 12 \ba\otimes\ba,\
 \ba\in \mathbb{R}^2,
\end{equation} we state the classical formulation of the problem:

We look for $\{\bu,u\}$ such that the system

\begin{align} \label{systfk} \left. \begin{array}{rl}
&\ddot{\bu}-  \Div (\mathfrak{C}_1+\mathfrak{C}_0)=\bF,\\[.5ex]
& \ddot{u} - a \lap \ddot{u} + b(e_1 \lap^2 \dot{u} + e_0 \lap^2 u)
-\mdiv \big( (\mathfrak{C}_1+\mathfrak{C}_0)  \nabla u \big) =f+p(u + g)
\end{array} \right\} \text{ on } Q,
\end{align}\\[.5ex]
holds, the boundary value conditions

\begin{align} \label{bvcfk}
 \left. \begin{array}{rl}
(\mathfrak{C}_1+\mathfrak{C}_0)\bn \cdot \bn = \tilde q(\ti{u_n} ),\ 
(\mathfrak{C}_1+\mathfrak{C}_0)\bn\cdot \btau = 0,\ u= u^{(0)} \\
e_1 (\lap \dot u+(1-\nu_1) B\dot u) + e_0 (\lap  u+(1-\nu_0) B u)= 0
\end{array} \right\} \text{ on } S
\end{align}
with
\begin{align*}
\ti{u_n} \equiv \bu \cdot \bn,\
B w=2n_1n_2\partial_{12} w-n_1^2\partial_{22}w
-n_2^2\partial_{11}w\\
\end{align*}
are satisfied, and the initial conditions

\begin{equation}  \label{initfk}
  \bu(0, \cdot)= \bu^{(0)},\  \dot{\bu}(0, \cdot) = \bu^{(1)}, \
		u(0, \cdot)= u^{(0)},\  \dot{u}(0, \cdot) = u^{(1)}  \text{ on } \Om
\end{equation}
are valid. Both functions $p$ and $-\tilde q$ are assumed to satisfy all the conditions to $p$
in Sec.~2. 
We remark that we assume our plate to be simply supported, because
it does not seem physically reasonable to consider the clamped plate with
the possible limited interpenetration on the boundary.   The constants $a,b,e_0,e_1$ are positive,
the nonnegative function $g$ 
is again the gap function. Of course, we can introduce another gap function to the 
boundary contact, but it seems to have a little use in practical applications. Let us remark that such
defined problem describes the behaviour of a cover of a fully recessed stack. 

For $z,y \in L_2(I; H^2(\Om))$ we define the following bilinear forms:
\begin{equation}\label{bilform}
  A_i : (z,y) \mapsto  b  e_i\big(\partial_{kk} z \partial_{kk} y +
  \nu_i (\partial_{11}z \partial_{22}y + \partial_{22}z \partial_{11} y)
  + 2(1-\nu_i)\partial_{12}z \partial_{12}y \big), i = 0,1.
\end{equation}
Then our problem has the following variational formulation: \\
{\em Look for $\{\bu,u\} \in \bH^1(Q) \times \big( L_2(I; H^2(\Om)) \cap L_2(I; \oH^1(\Om)) \big)$ such that $\dot{\bu}
\in L_2(I; \bH^1(\Om) ),\ \dot{u}\in L_2(I; H^2(\Om))$,
  and the system}
\begin{align}\begin{split}\label{vffk}
    &\IQ{} \big( ( \mathfrak{C}_1+\mathfrak{C}_0) \varepsilon(\by)
   -\dot \bu\cdot  \dot \by \big)\,dx\,dt+\int_\Om \big( (\dot \bu\cdot \by)(T,.)
    - \bu^1\cdot \by(0,.) \big)  \,dx \\
		& \kern 5em = \IQ{} \bF \cdot \by \,dx \,dt -  \int_{S} q(\ti{u_n} ) y_n\, dx_s \, dt,
   \\[.5em]
 & \IQ{ }\big(A_1( \dot{u}, z) +A_0(u,z) +[(\mathfrak{C}_1+\mathfrak{C}_0)\nabla
u] \cdot \nabla z)  -\dot u \dot z -a\nabla \dot u \cdot \nabla \dot z  \big)\,dx \,dt + \\
 &  \int_\Om \big( (\dot u z + a\nabla \dot u \cdot \nabla z )(T,\cdot) -
   u^{(1)} z(0,\cdot)  - a\nabla u^{(1)} \cdot \nabla z(0,\cdot)\big)\,dx = \IQ{ } (f+ p(u + g)) z \,dx \,dt
\end{split}
\end{align}
{\em is satisfied for every $\{\by, z\} \in \bY$} with
\begin{align} \label{testspacefull} \begin{split}
\bY \equiv \bY_0 \cap \bY_d  \text{ with } & \bY_0 \equiv \big\{ L_2(I; \bH^1(\Om)) \times \big(L_2(I; \oH^1(\Om)) \cap L_2(I; H^2(\Om))\big)  \big\}.\\
& \bY_d \equiv \{\bz \in \bY_0; \dot \bz \in \bL_2(Q)\}.
\end{split}
\end{align}

The approximate problems are defined as usually by replacing $p, q$ by $p_k, q_k$, respectively, and 
keeping acceleration terms in such modified system \eqref{vffk} in their original form. Since this problem is remarkably
more complex that all previous ones and leads to more complex formulae, we shall denote the solution of this problem 
also by $\{\bu, u\}$. Similarly to the previous problems
this approximate problem is again solved with the help of the Galerkin approximation. Since there is no substantial difference
between it and the penalized problem treated in \cite{full}, we postpone the readers to that paper for details. To derive
{\em a priori} estimates for the solutions of the approximate problem we put $\chi_{Q_s}\{\dot \bu, \dot u -\dot u^{(0)} \}$
for $s \in (0,T]$ as a test function of the appropriate variant of \eqref{vffk}. We obtain after the
integration and the summation
\begin{align} \begin{split}  \label{penint}
   &\IQ{s} \bigg(\frac{1}{2}  \partial_t\big(\dot{u}^2+a\vert \nabla
   \dot{u}\vert^2+\vert\dot{\bu}\vert^2
   +\rs{C}_0 \big( \varepsilon(\bu)+\Psi(\nabla u)\big) \cdot 
	(\varepsilon(\bu)+\Psi(\nabla u))
	+ A_0(u,u)) \big)\\
 &  + A_1( \dot{u}, \dot{u})+
  \rs{C}_1 \big(\varepsilon(\dot \bu)+\partial_t\Psi(\nabla u)\big)
  \cdot (\varepsilon(\dot \bu)+\partial_t \Psi(\nabla u)] + \partial_t  P_k(u + g)
    \bigg) dx\, dt   \\
		&  +  \int_{S}\  \partial_t \tilde Q_k(\ti u_n ) dx_s \,dt 
  =  \IQ{s}\left(\bF\cdot \dot \bu + f \dot{w}\right)\, dx \,dt. + R(u^{(0)}),
\end{split}
\end{align}
where $\tilde Q: r \mapsto \int_\infty^r q(\zeta) d\zeta$ and
$R(u^{(0)} )$ sums up all the terms containing $u^{(0)}$ or its derivatives.
Using the coercivity of the form $A_i$ and the form of the operators
$\rs{C}_i$  we obtain the estimate
\begin{align}
&  \| \dot u(s)\|_{H^1(\Om)}^2+\| u(s)\|_{H^2(\Om)}^2+ \|
\dot{\bu}(s)\|_{\bL_2(\Om)}^2 +\|\varepsilon(\bu)(s)+\Psi (\nabla
u)(s)\|_{L_2(\Om; \ST)}^2\nonumber\\
&  +\| \dot u\|_{L_2(I_s; H^2(\Om))}^2+\|\varepsilon(\dot
\bu)+\partial_t \Psi(\nabla u)\|_{L_2(Q_s; \ST)}^2+ \| P_k(u(s) +g)\|_{L_1(Q)} +
\label{penest}\\
&  \|\tilde Q_k (\ti u_n(s) )\|_{L_1(S)} \le C(\bu^{(0)},\bu^{(1)},u^{(0)} ,u^{(1)},\bF,f)\ 
\forall s\in (0,T].\nonumber
\end{align}
Applying the continuous imbedding $H^2(\Om)\hookrightarrow
W_4^1(\Om)$ we obtain the estimate
$$\|\Psi(\nabla u)(s)\|_{L_2(\Om; \ST)}+\left\|\partial_t \Psi(\nabla
u) \right\|_{L_2(I_s; L_2(\Om; \ST))}\le C(\bu^{(0)},\bu^{(1)},u^{(0)} ,u^{(1)},\bF,f)\
\forall s\in (0,T]$$ which implies
$$\|\varepsilon(\bu)(s)\|_{L_2(\Om; \ST)}+\|\dot \varepsilon(\bu)
\|_{L_2(I_s; L_2(\Om; \ST))}\le C(\bu^{(0)},\bu^{(1)},u^{(0)} ,u^{(1)},\bF,f)\ \forall
s\in (0,T].$$ Using the coerciveness of strains (see e.g. \cite{ejk}, Thm
1.2.3) we obtain
$$\| \bu(s)\|_{\bH^1(\Om)}+\| \dot \bu
\|_{L_2(I_s; \bH^1(\Om))}\le C(\bu^{(0)},\bu^{(1)},u^{(0)} ,u^{(1)},\bF,f) ) \ \forall
s\in (0,T]$$ 
which together with \eqref{penest} implies the {\em a priori} estimate
\begin{align}\begin{split}
&\| \dot \bu\|_{L_\infty(I; \bL_2(\Om))}+ \| \dot \bu \|_{L_2(I;  \bH^1(\Om))}+\|
\bu\|_{L_\infty(I; \bH^1(\Om))}+
 \| \dot u\|_{L_\infty(I; H^2(\Om))}+\| \dot u\|_{L_2(I; H^2(\Om))} \\
  &+ \| u\|_{L_\infty(I; H^2(\Om))}
+ \| P_k(u + g)\|_{L_\infty(I; L_1(Q))} + \|\tilde Q_k (\ti u_n ) \|_{L_\infty(I; L_1(S)]} \le \\
&\kern 5em   C(\bu^{(0)},\bu^{(1)},u^{(0)} ,u^{(1)},\bF,f) \ \label{aefull}
\end{split} \end{align}

Since there is no substantial difference between the proof of the solvability of
our approximate problem and that of the penalized problem treated in \cite{full}, we postpone the readers to that paper for details.
Via the standard method it is proved that such a solution is unique.

 As in all previous problems the main task is to perform the limit process $k \to \infty$ for which the $k$--independent estimates 
of the acceleration terms are needed. To estimate $\ddot \bu_k \in L_2(I; \bH^{-1}(\Om))$ we put an arbitrary $\bw \in L_2(I; \obH^1(\Om))$
in the approximate variant of \eqref{vffk} and use \eqref{aefull}.   
To get the estimate $\ddot u \in L_1(I; H^2(\Om)^* )$ we must assume \eqref{assimp}
which yields the uniform estimate for $\|p_k(u_k + g)\|_{L_1(Q)}$  as in Sec.~2. Then we are in the same situation as in Example 3, the Aubin Lemma gives us the crucial strong
$L_2(Q)$--convergence of all components of velocities. As earlier this leads to the upper semicontinuity of $\langle p_k(u_k) u_k\rangle_Q$ and $\left\langle \tilde q_k\!\left(
\ti{(u_k)_n}\right)  \ti{(u_k)_n} \right\rangle_S$ and finally to the fact that $p_k(u_k + g) \wto p(u + g)$ in $\cX_1^*$ as in Sec.~2 and $\tilde q_k\!\left(\ti{(u_k)_n}\right) \wto \tilde q(\ti{u_n})$
in $L_\infty^*(S)$ and we are done. The existence of solutions to \eqref{vffk} is proved.\\

Similarly to the previous sections we can formulate the full von K\'arm\'an system with the singular memory replacing all ``short memory'' terms in 
\eqref{systfk} and \eqref{vffk} by the corresponding singular memory ones. As in the previous cases under the assumption \eqref{smallmem} it is possible 
to pass from the appropriate approximate problem to the original one in such a way that the crucial strong convergence of velocities holds which leads
to the same conclusion as mentioned in the previous paragraph. We allow ourselves to leave this case to kind readers as an exercise.    

\section{Relation to the Signorini contact}

In this section we shall prove that for a sequence of the problems with the thickness of the interpenetration 
\begin{equation}
\gamma_\ell \nearrow 0,\  \ell \in \N,
\end{equation}
there is a subsequence of their solutions called $u_\ell$ tending to a limit $u$ which is a solution of a problem without interpenetration, i.e. 
of the appropriate Signorini version of the problem. Since there is the well-known generic nonuniqueness of the solutions to the dynamic
contact of the Signorini type related with the lack of  information about the amount of the energy conservation in the contact and thus about the
development of the solution after the contact, probably nothing more can be proved in general.

The common feature of the problems treated in the previous sections is that the estimates performed there as $k$--independent are also
$\gamma$ independent. Hence if we have a sequence $u_\ell$ tending weakly or weakly$^*$ to $u$ in the spaces for which the a priori 
and dual estimates have been derived, we have the strong $L_2$ convergence $\dot u_\ell \to \dot u$. Obviously for the full von K\'am\'an
system $\dot \bu_\ell \to \dot \bu$ in $\bL_2(Q)$ holds as well (cf. \cite{ejs}). Since $u_\ell  + g \ge \gamma_\ell$ a.e. in $Q$, we have $u + g  \ge 0$
a.e. there. Moreover for the full von K\'arm\'an system we get similarly  $\ti{u_n} \le 0$. We define
\begin{align} \begin{split} \label{cones}
  & \cK \equiv \{v \in \cX_0; v \ge -g \text{ a.e. in } Q\} \text{ for problems in Sections 2, 3,} \\
	& \cK \equiv  \{ \{v,\bom\} \in (u_0 + \oH^1(Q))\times \bX(Q); v \ge -g \text{ a.e. in } Q \}   \text{ for Reissner-Mindlin plates,}\\
  & \cK \equiv \{\{\bw,v\} \in  \bY_0; \bw_n \le 0 \text{ a.e. in } S,  v \ge -g \text{ a.e. in } Q\} \text{ for full von K\'arm\'an system,}
\end{split} \end{align}
if we define $\bX(Q)= L_2(I; \bX(\Om))$,
cf. \eqref{wcfc}, \eqref{vfrm}, and \eqref{testspacefull}. Obviously in all cases $p(v+g) =0$ if $v$ is (possibly a component) from $\cK$ and, moreover,
$q(w_n) = 0$ in the last case. Denoting $\Theta \equiv \lim_{\ell \to \infty} \langle p_\ell (u_\ell + g), u_\ell \rangle_Q$ and $\vartheta \equiv 
\lim_{\ell \to \infty} p_\ell(u_\ell + g)$, the monotonicity of $p_\ell$ used  for the couple $\{u_\ell,u\}$ yields $\Theta \ge \vartheta u$. On the other hand,
we can derive from \eqref{wcfc} (the solution there must be denoted by $u_\ell$) with $v = u_\ell -y$, $y \in \cK$ the opposite inequality,
because in general for $y \in \cK$ the lower semicontinuity of $\cA$ and $\cB$ in the limit process $\ell \to \infty$ yields 
\begin{align} \begin{split} \label{Sig}
 &-\langle \dot u, \dot y -\dot u \rangle_Q +  \langle \cA u, y-u \rangle_Q +  \langle \cB u, y-u \rangle_Q +  \langle \cE u, y-u \rangle _Q 
   +  \langle \vartheta, y \rangle_Q - \Theta\\
 & + \langle \dot u(T, \cdot), (y-u)(T, \cdot)\rangle_\Om \ge \langle f, y-u \rangle_Q +\langle u_1,y(0,\cdot)-u_1 \rangle_\Om
\end{split} \end{align}
and putting $y=u$ we are done. Hence \eqref{Sig} holds just without the terms with $\vartheta$ and $\Theta$ and this is the exact
formulation of the corresponding Signorini problem with $u$ being its solution. This pattern can be exactly followed also in all other cases, 
because their variational formulations contain only some lower semicontinuous parts, strongly converging terms and linear terms
for which the weak convergence is sufficient, hence we prove everytimes $\Theta = 
\langle \vartheta , u \rangle_{Q}$ and then we can see that the resulting limit variational inequality is the variational formulation
of the corresponding Signorini problem indeed. Of course, for Reissner-Mindlin plates we keep the second equation of \eqref{vfrm} in the original
form observing that the convergences which remain weak there are sufficient.

We only mention the full von K\'arm\'an system more in detail. We denote 
$$\Lambda \equiv \lim_{\ell \to \infty} \left\langle q\big(\ti{(u_\ell)_n} \big),  \ti{(u_\ell)_n} \right\rangle_{S} 
  \text{ and  } \lambda  \equiv \lim_{\ell \to \infty} q\big(\ti{(u_\ell)_n} \big).$$
and derive immediately that $\Lambda \ge \langle \lambda, \ti{u_n} \rangle_S$.  Denoting the solution of
\eqref{vffk} by $\{ \bu_\ell, u_\ell\}$ we put $\{\bv - \bu_\ell, w-u_\ell\}$ as a test function in \eqref{vffk} for an arbitrary
$\{\bv,w\} \in \cK$. Then we perform the limit process $\ell \to \infty$, denote by $\{\bu,u\}$ the limit of $\{\bu_\ell, u_\ell\}$
and put $\{\bw,v\} = \{\bu, u\}$ as a test function into the resulting inequality. Thus we find $\Lambda = \langle \lambda, \ti{u_n} \rangle_S$
from the first row of it while $\Theta = \langle \vartheta, u \rangle_Q$. From this we get that the resulting inequality is in fact the variational 
formulation of the Signorini contact for the full von K\'arm\'an system and $\{\bu, u\}$ is its solution.

\section{Conclusion}

Existence of solutions has been proved for the dynamic contact problems with limited interpenetration for viscoelastic variants of several classical models of plates.
These results are now available for technical practice. Probably the most challenging task for the application is the determination  of the function $p$ which 
describes the interpenetration. Performing some sensitivity analysis with respect of its choice may help here.


{\footnotesize \begin{thebibliography}{99}
\bibitem{karmve} I.~Bock and J.~Jaru\v{s}ek. Unilateral dynamic contact
of viscoelastic von K\'arm\'an plates {\em
Adv.~Math. Sci. Appl.}~{\bf 16}\,(1) (2006),
175--187.
\bibitem{karmsg} I.~Bock and J.~Jaru\v{s}ek, Unilateral dynamic contact
of von K\'arm\'an plates with singular memory. {\em Appl.~Math.}
{\bf 52}\,(6) (2007), 515-527.
\bibitem{full}  I.~Bock and J.~Jaru\v{s}ek. Dynamic contact problem for a bridge modeled
by a viscoelastic full von K\'arm\'an system. {\em Z. Angew. Math. Phys.} {\bf 61}\,(5) (2010),
865--876.
\bibitem{rm}  I.~Bock and J.~Jaru\v{s}ek. Unilateral dynamic contact problem for viscoelastic
Reissner-Mindlin plates. {\em Nonlin.~Anal., Theory Meth. Appl.} {\bf 74}\,(12), (2011), 4192--4202.
\bibitem{bz} J.M.~Borwein,~J and Q.J.~Zhu. {\em Techniques of Variational Analysis.} Springer, New York: 2005.
\bibitem{ejk} C. Eck, J. Jaru\v sek and M. Krbec: {\em Unilateral  Contact
  Problems: Variational Methods and Existence Theorems}, Monographs and Textbooks in Pure and
  Applied Mathematics {\bf 270}, Chapman/CRC Press, Providence R.I. etc., 2005. 
\bibitem{ejs} C. Eck, J. Jaru\v{s}ek and J. Star\'a. Normal Compliance Contact Models
  with Finite Interpenetration. {\em Arch. Ration. Mech. Anal.} {\bf 208}\,(1) (2013), 25--57.
\bibitem{j} J. Jaru\v{s}ek. Static semicoercive normal compliance contact problem with limited
  interpenetration. {\em Zeitschr. Angew. Math. Phys.} {\bf 66} (20150, 2161--2172.
\bibitem{js} J. Jaru\v{s}ek  and J. Star\'a. Solvability of a rational contact model
with limited interpenetration in viscoelastodynamics. {\em Math. Mech. Solids}  {\bf 23}\,(7) (2018), 1040--1048. 
\bibitem{ks} H.~Koch and A.~Stachel. Global existence of
classical solutions to the dynamic von K\'arm\'an equations. {\em
Math. Meth. Appl. Sci.} {\bf 16} (1993), 581--586.
\bibitem{Lagn}  J.E.~Lagnese. {\em Boundary Stabilization
of Thin Plates.} SIAM, Philadelphia,~PA,~1989.
\bibitem{S1} A.~Signorini. Sopra alcune questioni di statica dei sistemi continui. {\em Ann Scuola Norm-Sup.} {\bf 2} (1933): 231--251.
\bibitem{S2} A.~Signorini. Questioni di elasticit\`a non linearizzata e semilinearizzata, {\it Rend. Mat.}
  {\bf 18} (1959), 95-139.
\end{thebibliography}
}
\end{document}